\documentclass[epj,referee]{svjour}
\usepackage{graphicx}
\begin{document}
\title{Carbonyl sulphide under strong laser field: time-dependent density functional theory}
\titlerunning{Carbonyl sulphide under strong laser field}

\author{G. Bilalbegovi\'c}
\institute{Department of Physics, Faculty of Science, University of Zagreb,\\
Bijeni{\v c}ka 32, 10000 Zagreb, Croatia}
\date{To be published in The European Physical Journal D}
\abstract{
The first 52 fs of a time evolution of the electron density in  OCS after an interaction with an intense sub 10 fs
laser pulse are studied using the time-dependent density functional theory. The nuclear motion
in this linear trimer is simulated by the classical molecular dynamics method.
Laser fields of intensity $10^{13}$ W/cm$^2$ and $10^{15}$ W/cm$^2$ are used.
Details of the laser induced changes of the structure, as well as the ionization rate
are sensitive to the applied field intensity and its polarization.
It is found that under suitable conditions the OCS molecule bends soon after an interaction
with a laser pulse. A deviation from the linear geometry of up to
$23.6^\circ$ and charged ions of up to +3 are observed.
The time evolution of electric dipole moments and
the time-dependent electron localization function (ELF) are also studied.
 % end of PACS codes
} %end of abstract
\maketitle
\section{Introduction}
\label{intro}

An interaction between nanoparticles and strong laser fields is now an active area of research
\cite{Yamanouchi,Marangos,Marangos2,Corkum}.
Femtosecond pulsed lasers are able to produce high-intensity fields and they are used to initiate and investigate
various processes in nanoparticles \cite{Yamanouchi,Marangos,Marangos2}.
It is customary to call intensities above $10^{13}$ W/cm$^2$ as intense
laser fields, and time intervals below $10^{-13}$ s as ultra-short laser pulses.
Experimental and theoretical studies have been shown that
the multi-photon ionization mechanism
dominates a laser-molecule interaction  at intensities $ < 10^{14}$ W/cm$^2$
for laser frequencies in the near infrared part of the spectrum.
The tunnel and multi-electron dissociative ionizations
prevail for field intensities greater than $10^{14}$ W/cm$^2$.
Fields below $10^{13}$ W/cm$^2$ are not intense enough to induce the multiple ionization
of neutral molecules.
Research efforts are now oriented towards attosecond laser pulses and intensities of
$10^{24}$ W/cm$^2$ \cite{Corkum}.
When atoms, molecules, or clusters, accelerated to a high kinetic
energy, impinge on thin foils they sometimes undergo through the
process of Coulomb explosion. It is also
possible to achieve similar conditions if a nanoparticle collides with
an accelerated beam of  other particles, or interacts with intense laser
fields \cite{Frasinski,Dermota}. In these processes often occurs
the multiple ionization of a targeted nanoparticle. This nanoparticle
then explodes as a result of the strong Coulomb
repulsion. One or several charge centers may form in an initial
nanoparticle when the loss of electrons occurs under
laser light. Products of the Coulomb explosion are high-energy electrons
and ions, and under suitable conditions even X-rays. A behavior of
fragments in the Coulomb explosion sheds light on interactions
in an initial nanoparticle. The Coulomb explosion imaging is
successfully used to determine an equilibrium structure of
nanoparticles.

The carbonyl sulphide is an important material present in the Earth's atmosphere, volcanic gases, ice cores in Antarctica,
and the interstellar medium. It catalyzes
the formation of biological molecules and it has been suggested that OCS played an important role in the origin of life on our planet \cite{Leman}.
The charged clusters of carbonyl sulphide have been also studied
\cite{Surber,Goranka}.
In its ground state the OCS molecule is linear, but it is bent in the first excited state
\cite{Suzuki,Sugita}. In addition,
the OCS molecule consists of three different atoms. Therefore, in comparison with homonuclear trimers,
it is more prone to various modifications of the shape in  chemical processes and under strong fields.
It has been shown that the OCS molecule at high energies close to dissociation exhibits chaotic dynamics
\cite{Paskauskas}.
The Coulomb explosion of carbonyl sulphide has been investigated using ion-momentum imaging techniques
\cite{Sanderson}. The laser pulses of 55 fs, up to $2 \times 10^{15}$ W/cm$^2$ of intensity, and
wave length of $\lambda = 790$ nm have been used. The shape of the exploding OCS molecule has been determined.
Measured ion momentum maps have been compared to those obtained using the Monte Carlo simulation.
The ion momentum maps have been shown that  O$^{3+}$ and S$^{3+}$ exist along the laser polarization direction.
The central C$^{3+}$ exhibits a four-lobe structure which indicates that a non-linear geometry of OCS develops in the Coulomb explosion. A peak bond angle of 170$^{\circ}$
and bonds stretching have been measured. The O-C and C-S bonds doubled in an interaction with laser pulses.
In a recent work the stronger field of $\simeq 10^{16}$ W/cm$^2$
has been used to
investigate the multiple ionization of OCS \cite{Bryan}. The three dimensional covariance mapping technique has been applied.

The time-dependent density functional theory (TDDFT) \cite{Gross,Marques,Burke}
is a powerful technique for a description of excitations in nanoparticles.
It is possible to use this method to study
particle dynamics under strong laser pulses
\cite{Marques,Castro,Gonzalez,Alonso,Burnus,Suraud}. The time-dependent Schr{\"
o}dinger equation is replaced by the time-dependent Kohn-Sham one (atomic units are used)
\begin{equation}
i\frac{\partial}{\partial t} \phi_i(\vec{r},t) = \Bigl [-\frac{1}{2} \bigtriangledown ^2 + v_{KS}(\vec{r},t)\Bigr ]\phi_i (\vec{r},t).
\end{equation}
In this equation $v_{KS}(\vec{r},t)= v_H(\vec{r},t) + v_{xc}(\vec{r},t) + v_{ext}(\vec{r},t)$,
where
$v_H$ is the Hartree potential, $v_{xc}$  is the exchange-correlation potential, and  $v_{ext}$ is an external potential.
The particle density $n$ is calculated from
\begin{equation}
n({\vec r},t) = \sum _i^{N} \phi_i ^*(\vec{r},t)\phi_i (\vec{r},t),
\end{equation}
where $N$ is the number of occupied states.
The laser field enters in an external term of the Kohn-Sham
potential. In the classical dipole approximation the potential of the laser field is
\begin{equation}
V({\vec r},t) = I f(t) \sin ({\omega t}) \sum_{i=1}^N{{\vec r} \cdot {\vec \alpha}},
\end{equation}
where $I$ denotes the intensity of the field, $\omega$ is the frequency, and the ${\vec \alpha}$ is the polarization
of the light. The function $f(t)$ determines the shape of the laser pulse. The laser potential is a strong term and
it is necessary to solve the Kohn-Sham equations in a non-perturbative regime.

The Born-Oppenheimer approximation is often applied in the analysis
of the molecules. It assumes that the motion of nuclei and electrons
can be separated. Nuclei are much heavier and therefore  their
motion often can be neglected when electrons move. However, in some
physical situations the Born-Oppenheimer approximation breaks, and
it is important to treat the coupled dynamics of electrons and
nuclei. One example of a such coupled motion occurs for molecules in
strong laser fields. In the calculations presented here nuclei are
treated as classical point particles. The Newton equations for the
nuclei are solved for the force which is calculated  using the
Ehrenfest's theorem \cite{Octopus,Ullrich}. Electrons are described
by the quantum TDDFT. They are exposed to the Coulomb field of
classical nuclei and the laser field. This approach produces errors
when the molecule dissociates under a strong laser light. In the
theoretical analysis of the laser-molecule interaction it is
difficult to treat on the same level situations where the particles
in the molecule are bonded together and where nuclear densities are
delocalized. The classical treatment of nuclei brakes when a
probability for a dissociation of the molecule in the laser field is
high \cite{Burke,Kreibich}. The proper treatment of the coupling
between electronic and nuclear motion under strong laser light
should be carried out using the multi-component TDDFT
\cite{Gross2,Leeuwen}. In this theory both nuclei and electrons are
treated as quantum particles. However, this approach is still in
development. The multi-component TDDFT is computationally time
consuming and it has been performed only for small systems.

In this work mixed techniques of the
quantum mechanical time-dependent density functional theory and the classical
dynamics method  are used to
study the heteronuclear linear trimer
OCS after an irradiation by an intense and short laser pulse.
The results show a bending of the molecule and  an elongation of its bonds,
as well as ionizations, intensive energy exchanges and charge redistributions.
In the following the computational method is described in Sec. II.
Results and discussion are presented in Sec. III. A summary and conclusions
are given in Sec. IV.

\section{Computational method}
\label{sec:1}

The Coulomb explosion of the carbonyl sulphide is studied by
TDDFT method.
Calculations are performed using the Octopus  code \cite{Octopus}, where a real-space, pseudopotential
based treatment of
the TDDFT has been developed. Physical quantities are expanded using a mesh in the real space.
In addition, the ground state properties of OCS are also studied by the  pseudopotential and planewave based
Abinit code \cite{Abinit}.
The Troullier-Martins pseudopotentials \cite{Troullier} are used in both simulations, and
the OCS molecule is positioned
in the parallelepiped box with the side of $30$ a.u.
A Cartesian grid with the constant
mesh spacing of 0.2  \AA  \,
is applied in simulations based on the Octopus code.
The ground-state of OCS obtained using the density functional theory
is perturbed by the laser pulse. The time-dependent Kohn-Sham equations are integrated
using the time step of $\Delta t = 0.002$ $\hbar/eV$ $= 1.32 \times 10^{-18}$ s.
The time evolution of up to 52.8 fs is followed. This is a long time for the TDDFT method.
The approximated enforced time-reversal symmetry algorithm is applied  to treat the evolution operator in
the time-dependent Kohn-Sham problem \cite{Rubio}.
The adiabatic local density approximation exchange functional in the parametrization of
Perdew and Wang  is chosen \cite{Perdew}.  In this work
the velocity Verlet algorithm is used to follow the classical time evolution of atomic cores.
The OCS molecule is oriented along the z axis and
exposed to the linearly polarized laser light  along the x, y, or  z axis.
Both O-C-S and S-C-O orientations towards an incoming laser pulse
are studied for the light polarized along the molecule axis.
The absorbing conditions are used at the boundaries of the box \cite{Fevens}.

Electronic excitation spectra of carbonyl sulphide have been
measured from 5 to 360 eV using dipole spectroscopy
\cite{Brion,MPI}. The first absorption peak is rather broad and its
maximum is at 7-8 eV. Using TDDFT it is possible to calculate the
absorption spectrum as $ \sigma(\omega) \sim |d(\omega)|^2, $ where
$d(\omega)$ is the Fourier transform of the induced dipole moment
$d(t)=\int{ z n (\vec {r}, t) d^3r}$, and $n (\vec {r}, t) =
|\psi(\vec {r},t)|^2$ is an electron density. The absorption
spectrum of OCS, calculated using the Octopus code, shows that the
first peak is at 7 eV. This is in a good agreement with experiments
\cite{Brion,MPI}. Simulations of the laser-molecule interaction are
done at the nonresonant frequency of 2.5 eV, i.e., below the
first absorption peak.
Experimental studies of the
Coulomb explosion in OCS  have been performed at the nonresonant frequency of
$\lambda = 790$ nm $= 1.57$ eV \cite{Sanderson,Bryan}.
The length gauge, pulses with a cosine envelope, a total length of 6.6 fs,
and maximal
intensities of
$I_1 = 1.3272 \times 10^{13}$ W/cm$^2$ and $I_2 = 1.3272 \times 10^{15}$ W/cm$^2$
are applied in these simulations.

\section{Results and discussion}
\label{sec:3}

The ground state density functional theory calculation of the OCS
molecule produces a good agreement with experimental results. The
molecule is linear, the S-C distance is 1.548 $\AA$, whereas the O-C
one is 1.148 $\AA$. The measured values of these distances are 1.561
$\AA$ and 1.156 $\AA$, respectively \cite{Sanov}. The results for
changes of the geometry  when the
laser field is polarized perpendicularly to the initial  molecule
axis are shown in Table I.  The  C$-$S
distance starts to change at 3.96 fs, the O$-$C distance at 9.24 fs,
and the angle $\angle$ (O$-$C$-$S) at 14.52 fs, for the laser intensity $I_1$.
The O$-$C distance
first increases up to 1.209 $\AA$ (at 19.14 fs), and then decreases. For
example, the O$-$C distance is 1.164 $\AA$ at 26.40 fs. The  C$-$S
distance increases up to 1.689 $\AA$ at 31.68 fs and then decreases. The
results for the intensity $I_2= 1.3272 \times 10^{15}$ W/cm$^2$
also show that the  distances do not change monotonically in time.
The distances between atoms
when the laser field is
polarized along the initial molecule axis are  presented in Table II.
There the molecule is linear. In this Table the
results for two orientations of the molecule towards a laser pulse
are shown. It is calculated that data for these two
orientations  differ only quantitatively.
The results for the S-C-O orientation are presented in examples for an
interaction with the z-polarized laser light in Figs. 2-5. Figure 1
shows an example of the change of
the O$-$C and C$-$S atomic distances during the
52.8 fs of the time evolution.
Similar nonmonotonic changes of
the internuclear distance have been recently measured and calculated
in the D$_2^-$ molecule study by ultrashort laser fields on a time
scale of 3 ps \cite{Feuerstein}.

In general, the angle between bonds
changes when the laser polarization is along directions which are
perpendicular to the molecular axis.
OCS preserves its linear geometry when the laser light
is polarized along the molecule axis.
However, bending of OCS induced by pulses polarized along the internuclear
axis may occur on a longer time scale, i.e., after 52.8 fs.
The observed maximum of the
angle between bonds is $23.6^\circ$ and it is calculated for the
x-polarized pulse of intensity $I_2= 1.3272 \times 10^{15}$
W/cm$^2$. As in experiments \cite{Sanderson,Bryan}, it is found
that the O$-$C bond is stronger than the C$-$S bond, and that both
bonds often strongly increase after an interaction with a laser
pulse. It is already known that atomic bonds are substantially
elongated in comparison with equilibrium ones in the processes of
an enhanced multiple ionization and the Coulomb explosion
\cite{Seideman}.

\begin{table*}
\caption{\label{tab:table1} The time evolution of the structure  for
laser intensities $I_1= 1.3272 \times 10^{13}$ W/cm$^2$ and $I_2=
1.3272 \times 10^{15}$ W/cm$^2$, using
the polarization direction perpendicular to the initial axis of the
molecule (x polarization). The distances d (here, in Table II and Fig. 1)
are geometrical separations between atoms
(including the cases where they are not chemically bonded).}
\begin{tabular}{ccccccc}
\hline
& $I_1$  &  &  &  $I_2$  \\
\hline
time [fs] & d(O$-$C)  &  d(C$-$S)  & $\angle$ (O$-$C$-$S)  & d(O$-$C)  &  d(C$-$S)  & $\angle$ (O$-$C$-$S)\\
\hline
2.64 & 1.148  & 1.548 & $180^{\circ}$ & 1.148 & 1.548 & $180^{\circ}$\\
7.92 & 1.148 & 1.560& $180^{\circ}$ & 1.152 & 1.572 & $180^{\circ}$\\
13.20 & 1.172 & 1.576& $180^{\circ}$ & 1.184 & 1.616 & $179.2^{\circ}$\\
18.48 & 1.204 & 1.604& $178.7^{\circ}$ & 1.207 & 1.688 & $170.8^{\circ}$\\
26.40 & 1.164& 1.668  & $166.9^{\circ}$ & 1.088 & 1.824 & $171.7^{\circ}$\\
31.68 & 1.049  & 1.689  & $171.7^{\circ}$ & 1.012& 1.865 & $176.9^{\circ}$  \\
\hline
\end{tabular}
\end{table*}

\begin{table*}
\caption{\label{tab:table2}
The time evolution of the structure for the laser intensity $I_2= 1.3272 \times 10^{15}$ W/cm$^2$
and a laser pulse polarized along the internuclear axis (z polarization).
The data for two orientations (S-C-O and O-C-S) of the carbonyl sulphide molecule in the laser field are presented.}
\begin{tabular}{ccccccc}
\hline
& $S-C-O$  &  &  &  $O-C-S$  \\
\hline
time [fs] & d(O$-$C)  &  d(C$-$S)  & $\angle$ (O$-$C$-$S) & d(O$-$C)  &  d(C$-$S)  & $\angle$ (O$-$C$-$S) \\
\hline
2.64 & 1.148  & 1.548 & $180^{\circ}$ & 1.148  & 1.552 & $180^{\circ}$\\
7.92 & 1.184  & 1.548 & $180^{\circ}$ & 1.192  & 1.568 & $180^{\circ}$\\
13.20 & 1.260 & 1.572 & $180^{\circ}$ & 1.272  & 1.600 & $180^{\circ}$\\
18.48 & 1.336 & 1.648 & $180^{\circ}$ & 1.288  & 1.660 & $180^{\circ}$\\
26.40 & 1.288 & 1.844 & $180^{\circ}$ & 1.260  & 1.784 & $180^{\circ}$\\
31.68 & 1.176 & 2.008 & $180^{\circ}$ & 1.252  & 1.880 & $180^{\circ}$\\
\hline
\end{tabular}
\end{table*}

\begin{figure}
\includegraphics*[scale=0.5]{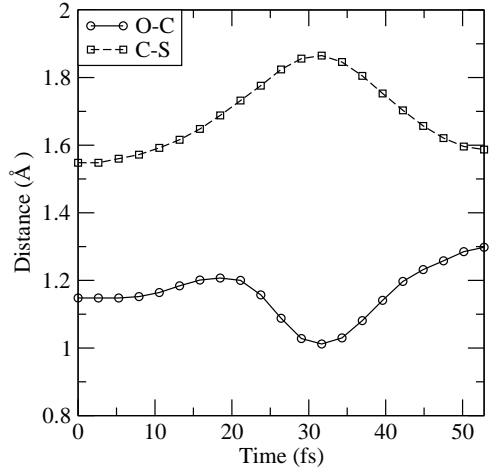}
\caption{The time evolution of atomic distances after an interaction with the laser light of
$I_2=1.3272 \times 10^{15}$ W/cm$^2$, x polarization.}
\label{fig:fig1}
\end{figure}

\begin{figure}
\includegraphics*[scale=0.7]{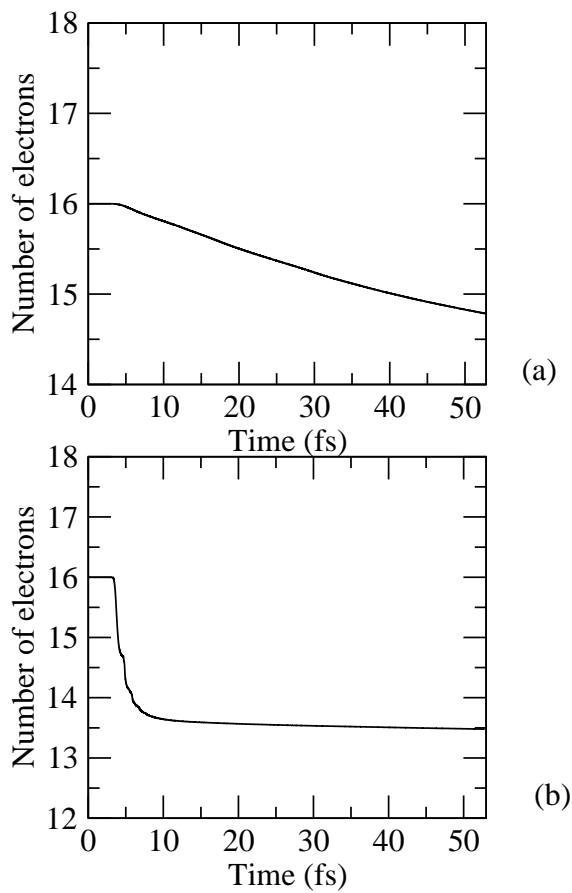}
\caption{The time evolution of the electronic charge after an interaction with the laser light:
(a) $I_2=1.3272 \times 10^{15}$ W/cm$^2$, x polarization,
(b) $I_2=1.3272 \times 10^{15}$ W/cm$^2$, z polarization.}
\label{fig:fig2}
\end{figure}

The OCS molecule ejects electrons after an interaction with a strong enough laser pulse.
Figure 2 shows how the electronic charge of this
16-valence electron molecule (as represented by the pseudopotential model)
changes when OCS is irradiated by a laser
pulse.
As shown in Figs. 2(b), charge states of up to +3  are observed under the field of
$\sim 10^{15}$ W/cm$^2$.
It is also found that for all laser light polarizations
the electronic charge in the box does not change (within the simulation time of 52.8 fs)
under the laser field of $I_1= 1.3272 \times 10^{13}$ W/cm$^2$.
The results show that the z-polarized laser light is more
efficient in removal of electrons from the OCS molecule. In this
simulation only the overall electronic charge of the molecule is
calculated by integrating the electron density in the simulation
box. However, in the experimental work and corresponding Monte Carlo
simulations several $(Q_1, Q_2, Q_3)$ fragmentation channels have
been proposed, where the labels are as in $(O^{Q_1+}, C^{Q_2+},
S^{Q_3+})$ \cite{Sanderson,Bryan}. For example, it has been found
that $(3, 3, 5)$ is a channel with the highest ionization under the
field of $\simeq 10^{16}$ W/cm$^2$.

\begin{figure}
\caption{The time evolution of the kinetic energy of OCS after an interaction with a laser pulse:
(a) $I_1=1.3272 \times 10^{13}$ W/cm$^2$,  z polarization,
(b) $I_2=1.3272 \times 10^{15}$ W/cm$^2$,  x polarization,
(c) $I_2=1.3272 \times 10^{15}$ W/cm$^2$,  z polarization.}
\label{fig:fig3}
\end{figure}

Several typical examples of the kinetic and potential energy are
presented in Figs. 3 and 4. It is well known that positively charged
ions form in the process of the laser-molecule interaction.
The Coulomb
repulsion of the ions and an initial binding energy of the
molecule are stored as the potential energy. The repulsive Coulomb
potential energy is strong and it is rapidly released as
a kinetic energy of the molecule fragments.
The results show that energies change after a time delay of at
least 2 fs. Regularly positioned  peaks are clearly visible in Fig.
3(b). The peaks are less pronounced under conditions of Fig. 3(a,c).
Therefore, the peak shapes and positions are intensity and polarization dependent.
These peaks in the time evolution of the kinetic energy and similar ones
for the potential energy (Fig. 4(a,b)) show that an intensive
redistribution of the energy occurs in the OCS molecule after an interaction
with a laser pulse.  Under conditions of Fig. 3(b) the laser field caused the molecule to bend. Then, the molecule starts to
vibrate and the result is an oscillatory exchange of kinetic and potential energy. The potential energy (Fig. 4(c)) is much larger
in an absolute value, and therefore the peaks are more visible in the kinetic energy (Fig. 3(b)).
It has been measured
for OCS perturbed by the laser light that the
kinetic energy release varies with the ionization channels
\cite{Bryan}. This study has been performed using laser
pulses focused to $\sim 10^{16}$ W/cm$^2$ and the kinetic energies
between 4.1 eV and 105 eV have been measured.   In the ground state of the molecule
its potential energy is negative. When the molecule is perturbed by
the laser light, the potential energy increases.
Figure 4 shows that the potential energy is always negative. Therefore,
the system preserves at least the
part of its initial binding energy.
The time evolution of the electric dipole moment $d_z(t)$ of the molecule is calculated
and presented in Fig. 5. When the light is polarized along the molecule axis,
the change of an electric dipole distribution in the first several fs is especially intensive
and approximately follows the oscillatory features of the laser pulse.

The electron localization function (ELF) has become an important tool
for a description of chemical bonds in molecules and solids
\cite{Becke,Savin}. The results of many studies have been shown that,
in comparison with electron densities, the ELFs better describe the
properties of localized and delocalized electrons in atoms and their
aggregates. Recently the time-dependent electron localization
function suitable for an analysis of the creation and
breaking of bonds has been introduced \cite{Burnus}. The three-dimensional
representations of  the time-dependent ELF isosurfaces for OCS are presented in Fig. 6.
The comparison of ELFs in the ground-state (Fig. 6(a)) and after 0.66 fs
(Fig. 6(b)) shows that the instability of the OCS
molecule first occurs at the sulfur atom (the upper left part of the ELF shown in Fig. 6(b)).
Figure 6(b) represents the point of the system time evolution when the distances are
d(O-C)= 1.148 \AA,
d(C-S)= 1.548 \AA.
Further analysis of the
time-dependent ELF confirms that the electron ejection begins around
the  S atom. Figure 6(c) shows several blobs
of a negative charge.
The corresponding geometry of the molecule shown in Fig. 6(c) is d(O-C)= 1.148 \AA,
d(C-S)= 1.560 \AA.
After 2.64 fs electrons are already
ejected from the molecule.

\begin{figure*}
\caption{The time evolution of the potential energy of OCS after an interaction with a laser pulse:
(a) $I_1=1.3272 \times 10^{13}$ W/cm$^2$, x polarization,
(b) $I_1=1.3272 \times 10^{13}$ W/cm$^2$, z polarization,
(c) $I_2=1.3272 \times 10^{15}$ W/cm$^2$, x polarization,
(d) $I_2=1.3272 \times 10^{15}$ W/cm$^2$, z polarization.
}
\label{fig:fig4}
\end{figure*}

\begin{figure*}
\caption{The time-dependent dipole moment (along the molecule axis) after an interaction with a laser pulse:
(a) $I_1=1.3272 \times 10^{13}$ W/cm$^2$, x polarization,
(b) $I_1=1.3272 \times 10^{13}$ W/cm$^2$, z polarization,
(c) $I_2=1.3272 \times 10^{15}$ W/cm$^2$, x polarization,
(d) $I_2=1.3272 \times 10^{15}$ W/cm$^2$, z polarization.}
\label{fig:fig5}
\end{figure*}

\begin{figure}
\caption{The time-dependent electron localization function  (ELF).
An initial ground-state structure is shown in (a). The same molecule after an
interaction with the x-polarized laser pulse of the  intensity $I_2=1.3272 \times 10^{15}$ W/cm$^2$:
(b) 0.66 fs,
(c) 2.64 fs.}
\label{fig:fig6}
\end{figure}

\section{Conclusions}
\label{sec:4}
An electron-ion dynamics of the carbonyl-sulphide
molecule exposed to the intense, sub 10 fs, linearly polarized laser pulse is
studied using calculations based on the numerical integration of
the equations developed within the time-dependent density functional
theory. The dynamics is followed for 52.8 fs. The results of this
simulation show that the OCS molecule bends during the ionization
process if the laser polarization direction is perpendicular to the
initial axis of the molecule. This molecule preserves its linear
morphology after 52.8 fs of the time evolution when it is oriented along the laser polarization
direction. An intensive redistribution of charges and energies,
and the stretching of bonds are observed.
These results are in a good agreement with experimental studies of the carbonyl sulphide molecule in
a strong laser field \cite{Sanderson,Bryan}.
However, experiments have been carried out on a much longer time scale.
TDDFT is an {\it ab initio} computational method
which is able to simulate processes in materials
at nanoscales induced by short and strong laser pulses.

\begin{acknowledgement}
This work has been carried out under the HR-MZOS project ``Electronic Properties of Surfaces and Nanostructures''.
I am grateful to the University Computing Center SRCE for their help and computer time.
\end{acknowledgement}

\bibliographystyle{epj}

\end{document}